\documentclass[prd,twocolumn,superscriptaddress,showpacs,preprintnumbers,amsmath,amssymb,nofootinbib,APS]{revtex4}

\usepackage{dcolumn}
\usepackage{bm}
\usepackage[latin1]{inputenc}
\usepackage[spanish,english]{babel}
\usepackage{amsfonts}
\usepackage{amssymb}
\usepackage{graphicx}

\newcommand{\be}{\begin{equation}}
\newcommand{\ee}{\end{equation}}
\newcommand{\bea}{\begin{eqnarray}}
\newcommand{\eea}{\end{eqnarray}}

\begin{document}

\title{{\bf Hawking radiation by Kerr black holes and  conformal symmetry}}
\author{Ivan Agullo}
\affiliation{ {\footnotesize Physics Department, University of
Wisconsin-Milwaukee, P.O.Box 413, Milwaukee, WI 53201 USA}}
\author{José Navarro-Salas}
\affiliation{ {\footnotesize Departamento de Física Teórica and
IFIC, Centro Mixto Universidad de Valencia-CSIC.
    Facultad de Física, Universidad de Valencia,
        Burjassot-46100, Valencia, Spain. }}

\author{Gonzalo J. Olmo}
\affiliation{\footnotesize Instituto de Estructura de la Materia,
CSIC, Serrano 121, 28006 Madrid, Spain}
\author{Leonard Parker}
\affiliation{ {\footnotesize Physics Department, University of
Wisconsin-Milwaukee, P.O.Box 413, Milwaukee, Wisconsin 53201 USA}}

\begin{abstract}
The exponential blueshift associated with the event horizon of a black hole makes conformal symmetry play a fundamental role in accounting for  its thermal properties. Using a derivation based on two-point functions, we show that the full spectrum of thermal radiation of scalar particles by Kerr black holes can be explicitly derived on the basis of a conformal symmetry arising in the  wave equation near the horizon. The simplicity of our approach emphasizes the depth of the connection between conformal symmetry and black hole radiance.

\end{abstract}


\maketitle

The prediction of thermal radiation by event horizons can be considered as one of the most remarkable outcomes of quantum field theory in curved spacetimes. This result and its implications, in particular those concerning black holes, constitute one of the best insights that we have at present about the features that a quantum theory of gravity should possess.  The thermal character of the radiation emitted by a black hole is linked to the presence of an event horizon. One characteristic of the event horizon is the existence of an unbounded (exponentially growing) gravitational blueshift that a particle with a given energy at infinity experiences as it approaches the horizon. This blueshift sweeps away any physical scale present in the field theory and makes conformal symmetry arise in a rather natural way. It has long been argued that the flux of the thermal radiation is deeply connected to anomalies related to the conformal symmetry arising near the horizon \cite{christensen-fulling, navarro-talavera, wilczek} (for an extensive account, see \cite{fabbri-navarro-salas}). Additionally, hints indicating that conformal symmetry suffices to yield  the {\em full spectrum} (not only the flux) of thermal radiation emitted by Schwarzschild black holes in four dimensions were obtained in Refs. \cite{agullo-navarro-salas-olmo-parkerPRL1,agullo-navarro-salas-olmo-parkerPRD1}. Higher-order moments of the Planck distribution were obtained in Ref. \cite{iso-morita-umetsu} through an involved analysis of higher-spin currents.

Furthermore, it has been argued
that the near-horizon conformal symmetry (which can emerge from different perspectives \cite{carlip,strominger, solodukhin}) is at the heart of the entropy of  black holes. Recently, the approach based on near-horizon asymptotic symmetries \cite{strominger} has been extended to  rotating (Kerr) black holes, suggesting a holographic duality between extremal and near-extremal Kerr black holes and a $2$-dimensional conformal field theory (CFT) \cite{kerrcft}. Additionally, the finite-temperature correlators of the dual CFT provide the amplitudes for scattering of particles off near-extremal Kerr black holes and therefore account for the phenomenon of superradiance \cite{superradiancekerrcft}. A further step has  more recently  appeared  in \cite{hiddencft}, where a (finite-dimensional) conformal symmetry has been shown to exist for the wave equation of a massless scalar field (in the so-called near region) for a  generic non-extremal Kerr black hole. This finite-dimensional $SO(2,2)$ symmetry accounts again for classical superradiance.

In view of the above results, it remains a challenge to show that the full thermal spectrum can be derived from conformal symmetry for a generic Kerr black hole. This is the main goal of our Letter.  
An essential point in our derivation is that, in the near-horizon region, the $SO(2,2)$ conformal symmetry of the classical matter wave equation is naturally enlarged to the conventional infinite-dimensional group of conformal transformations in two-dimensions.
This enlargement of the symmetry produces, at the quantum level, anomalous transformations of the vacuum state that  are responsible for the phenomenon of black hole thermal radiance. We show that the full spectrum of particles emitted by a rotating black hole can be obtained essentially as the Fourier transform of the two-point functions of primary fields of a $2$-dimensional CFT. The simplicity of our derivation illuminates in a new way the essential role of conformal symmetry in black hole radiance. This result may suggest new insight on the conjectured relation between Kerr geometry and a $2$-dimensional CFT.

Hawking's original derivation of black hole radiance \cite{hawk1} rests on the formalism of Bogolubov transformations in the context of gravitationally induced particle creation \cite{parker68}. In short, the derivation considers two vacuum states: The first, the {\it in} vacuum, coincides with the natural vacuum at early times before the star has begun to collapse, and the second, the {\it out} vacuum, coincides with the natural vacuum at late times long after it has collapsed to form a black hole as seen by a distant observer. The number of particles measured in the $i$th mode by an {\it out} observer when the field is in the {\it in} vacuum state, is obtained by evaluating the expectation value of the {\it out} number operator.
 This quantity can be computed as $\langle in|N^{out}_i |in\rangle=\sum_k |\beta_{ik}|^2$, where $\beta_{ik}$ are the Bogolubov coefficients relating the {\it in} and {\it out} field modes (for details see, for instance, \cite{parker-toms}). As explained in Refs. \cite{agullo-navarro-salas-olmo-parkerPRL1,agullo-navarro-salas-olmo-parkerPRD1}, this expectation value can be conveniently rewritten in terms of the two-point functions of the field as
\bea \label{eq:new-nord} \langle in|N^{out}_i |in\rangle = \int_{\Sigma} d\Sigma_1^{\mu} d\Sigma_2^{\nu} \left[f^{out}_i(x_1) {\buildrel\leftrightarrow\over{\partial}}_{\mu}\right]  \left[f^{out *}_i(x_2) {\buildrel\leftrightarrow\over{\partial}}_{\nu}\right] \\ \nonumber  \times \left[ \langle in| \Phi(x_1)  \Phi(x_2) |in\rangle - \langle out| \Phi(x_1)  \Phi(x_2) |out \rangle \right] \ ,\eea
where $f^{out}_i(x)$ represent the field modes defining the {\it out} vacuum state and $\Sigma$ is an arbitrary Cauchy hypersurface. The use of two-point functions here is particularly convenient to determine and take advantage of the symmetries present in the problem. We will use the above expression to compute the spectrum of scalar particles emitted by a Kerr black hole. First, we want to summarize some important features of $2$-dimensional CFT.

\noindent {\it $2$-dimensional CFT and two-point functions-} Let us consider a  $d$-dimensional spacetime. A conformal transformation of coordinates (see \cite{difrancesco} for details) is an invertible mapping $x \to x'$ which leaves the metric tensor invariant up to scale. The set of these transformation forms the conformal group $SO(d,2)$. The case $d=2$ requires special attention. In addition to the global transformations $SO(2,2)$, in $d=2$ the set of conformal transformations is enlarged to the infinite-dimensional group of local (not globally defined) transformations of the form $x^{+} \to x'^{+}(x^{+})$ and $x^{-} \to x'^{-}(x^{-})$, where $x^{\pm}= t \pm x$ are null coordinates. Additionally, in a $d=2$ conformal field theory there is a particular set of fields, called primary fields,  that under all (global and local) conformal transformations behave as $\label{2dCtransformation} \tilde\Phi (x^+,x^-) \to \tilde\Phi' (x'^+,x'^-)=\left(\frac{\partial x'^+}{\partial x^+}\right)^{-h^+} \left(\frac{\partial x'^-}{\partial x^-}\right)^{-h^-} \tilde\Phi (x^+,x^-)$, where $h^{\pm}$ are the so-called conformal weights. Typical examples of primary fields are the derivatives $\tilde\Phi\equiv  \partial_{\pm} \Phi $ of a 2-dimensional massless scalar field $\Phi$. For example,  we have $h^+=1$ and $h^-= 0$ for $\partial_+\Phi$ and the opposite weights for $\partial_-\Phi$.
Conformal invariance leads, in addition, to the following transformation law of the two-point function (for illustrative purposes, we will consider, for instance, $\partial_+\Phi$)
\bea \label{2deq:def} & & \langle 0|\partial_{+}\Phi(x_1)
\partial_{+}\Phi(x_2)|0 \rangle = \\ \nonumber & & \left(\frac{dx'^{+}}{dx^{+}}\right)(x'^{+}_1) \left(\frac{dx'^{+}}{dx^{+}}\right)(x'^{+}_2)
\langle 0|\partial_{+}\Phi(x'_1)
\partial_{+}\Phi(x'_2) |0\rangle   \ ,\eea
where $|0\rangle$ is the vacuum state of the theory. This requirement fixes the form of the two-point function to be
\be
\label{2dcorrelations}\langle 0|
\partial_{+}\Phi(x_1)
\partial_{+}\Phi(x_2) |0 \rangle = -\frac{1}{4\pi}\frac{1}{(x^{+}_1 -
x^{+}_2)^{2}}  \ . \ee
The vacuum state $|0\rangle$ is invariant under the global conformal group $SO(2,2)$. However, it is not invariant under local conformal transformations. This can be explicitly seen by computing  $\langle in|N^{out}_i |in\rangle$, where here the {\it in} and {\it out} sets of modes defining the $|in\rangle\equiv|0\rangle$ and $|out\rangle$ vacuum states are related by a conformal transformation. We can use expression (\ref{eq:new-nord}) to evaluate this expectation value. Integrating by parts in (\ref{eq:new-nord}) and taking into account that the field modes $f_i^{out}$ vanish at spacelike infinity, one finds that the two-point functions of the primary field $\partial_+\Phi$ emerge as the relevant ones. 
Relation (\ref{2deq:def}) can be then used to express the integral in (\ref{eq:new-nord}), for instance, in coordinates $x'$. Then, the kernel of the integral, given by the difference of two-point functions, reads 
\be \label{2d2pt}   -\frac{1}{4\pi}\frac{ \frac{dx'^{+}}{dx^{+}}(x'^{+}_1) \frac{dx'^{+}}{dx^{+}}(x'^{+}_2) }
{(x^{+}(x'^{+}_1) -
x^{+} (x'^{+}_2))^{2}}  + \frac{1}{4\pi}\frac{1}{(x'^{+}_1 -
x'^{+}_2)^{2}} \ . \ee
(A similar analysis holds for the opposite chiral sector.)
The important point here is that the elements of $SO(2,2)$ are the {\it only} transformations producing a vanishing result for this expression. This indicates the invariance of the vacuum state under global conformal transformations. However, the remaining (local) conformal transformations of the form  $x^{\pm} \to x'^{\pm}(x^{\pm})$ produce a nonvanishing value for expression (\ref{2d2pt}). The expectation value $\langle in|N^{out}_i |in\rangle$ is then nonvanishing, showing the nonequivalence of the $|in\rangle$ and $|out\rangle$ vacua and the phenomena of particle production. This is indeed a manifestation of the so-called Virasoro anomaly, which is at the root of the radiation of particles by black holes \cite{christensen-fulling, navarro-talavera, fabbri-navarro-salas,wilczek}.

\noindent {\it Scalar wave equation in Kerr geometry and $2$-dimensional conformal symmetry-} Let us consider the late time stages of the spacetime produced by the collapse of a rotating star, when the geometry is described by the stationary Kerr line element. In Boyer-Lindquist coordinates, it reads

\bea \label{kerr} ds^2 &=&\frac{\Delta}{\rho^2} (dt-a \sin^2{\theta} d\phi)^2-\frac{\sin^2{\theta}}{\rho^2} [(r^2+a^2) d\phi-a dt]^2 \nonumber  \\ &-&\frac{\rho^2}{\Delta} dr^2-\rho^2 d\theta^2 \ , \eea
where $\Delta=(r-r_+)(r-r_-)$, $\rho^2=r^2+a^2 \cos^2{\theta}$ and $r_{\pm}= M\pm (M^2-a^2)^{1/2}$. The parameters $M$ and $a$ represent the mass and the angular momentum per unit mass of the black hole respectively. We will assume that $a^2<M^2$. The event horizon of the Kerr black hole is located at $r=r_+$. The line element (\ref{kerr}) is stationary and axisymmetric, with $\partial^{\mu}_t$ and $\partial^{\mu}_{\phi}$ the corresponding Killing vector fields. One of the most interesting properties of Kerr geometry is the existence of a region outside the event horizon where the vector field $\partial^{\mu}_t$ becomes spacelike. The combination $\chi^{\mu}=\partial^{\mu}_t+\Omega_H \partial^{\mu}_\phi$, with $\Omega_H=a/(r_+^2+a^2)$, defines here the Killing vector field that generates the horizon. It is null at $r=r_+$ and timelike for $r>r_+$ in the near-horizon region $r-r_+\ll M$. It is then convenient to define the coordinates $\tilde{t}=t$ and $\tilde{\phi}=\phi-\Omega_H t$. We have then $\partial^{\mu}_{\tilde t}=\chi^{\mu}$. It is also appropriate to define the radial coordinate $r_*$ such that $dr_*/dr=(r^2+a^2)/\Delta$.

Let us now consider a massive scalar field $\Phi(x)$ propagating in the Kerr geometry  (\ref{kerr}). The Klein-Gordon equation $(\Box +\mu^2)\Phi(x)=0$ admits a full separability of variables (see, for instance, \cite{frolov-novikov} for details)
\be \Phi(t,r,\theta,\phi)=\sum_{A,m}  \frac{\Phi_{A,m}(r,t)}{\sqrt{r^2+a^2}} Z_{A,m}(\theta,\phi)\ , \ee
where $\Phi_{A,m}(r,t)=\int dw \Phi_{A,m}(r,w) e^{-i w t}$. The solutions of the angular equation, $Z_{A,m}(\theta,\phi)= (1/\sqrt{2 \pi})S_{A,m}(\theta)e^{i m \phi}$, are the so-called spheroidal harmonics. The functions $S_{A,m}(\theta)$ are eigenfunctions of the spheroidal angular equation with eigenvalue $A$.  The spheroidal harmonics form an orthonormal set of angular functions characterized by $A$ and $m$, such that $\int d\phi d(\cos{ \theta} )Z_{A,m}(\theta,\phi) Z^*_{A',m'}(\theta,\phi)=\delta_{A,A'} \delta_{m,m'}$.

The radial equation takes the form $[d^2/dr_*^2 + V(r)]\Phi_{A,m}(r,w)=0$. The potential $V(r)$ has a simple form in the asymptotic regions. As $r\to \infty$, $V \to w^2-\mu^2$ and the $(t,r)$-part of the solutions behaves as in Minkowski spacetime $\sim e^{-i (w t\pm k r)}$, with $k^2=w^2-\mu^2$. In the region near the horizon ($r_*\to -\infty$) it is more convenient to employ coordinates $(\tilde{t},r_*,\theta, \tilde{\phi})$. The potential takes the simple form $V\sim \tilde{w}^2$ as $r_*\to -\infty$, where $\tilde{w}=w-\Omega_H m$. This is a consequence of the exponential blueshift intrinsic to the event horizon, which shifts away any physical scale (such as the mass of the field or the centrifugal potential barrier). The physics near the horizon then can be described by the infinite set of $(1+1)$-dimensional massless fields $\Phi_{A,m}(\tilde{t},r_*)$ propagating in the  $(\tilde{t}, r_*)$ plane. In $(\tilde{t},r_*,\theta, \tilde{\phi})$ coordinates, the $(\tilde{t}, r_*)$ part of the solutions to the wave equation is a linear combination of ingoing and outgoing modes of the form $(e^{-i \tilde{w} (\tilde{t} + r_*)}\ , e^{-i \tilde{w} (\tilde{t} - r_*)})=(e^{-i \tilde{w} v}\ , e^{-i \tilde{w} u}$), where $v\equiv\tilde{t}+r_*$ and $u\equiv\tilde{t}-r_*$ are null coordinates. These modes are positive frequency modes with respect to the Killing vector field $\chi^{\mu}=\partial_{\tilde{t}}^{\mu}$ for $\tilde{w}>0$. They are solutions of the wave equation $\partial_u\partial_v\Phi_{A,m}(u,v)=0$ which is manifestly invariant under the infinite-dimensional group of conformal transformation in two dimensions $u\to u'(u)$, $v\to v'(v)$. In the following, we will show how this conformal symmetry is at the heart of the phenomenon of thermal emission of particles by black holes.

\noindent {\it Thermal radiation by Kerr black holes and $2$-dimensional CFT-} The phenomenon of thermal radiation by a rotating black hole was first derived by Hawking \cite{hawk1}. As mentioned above, in the original derivation Hawking considered the state of the field given by the vacuum defined at early times, before the star begins to collapse.
However, as pointed out by Unruh \cite{unruh} (see also \cite{fredenhagen-haag90}), the result is equivalently obtained by substituting the state of the field by the vacuum defined by a freely falling observer crossing the horizon at the time when the surface of the collapsing star enters the horizon. In that way, the vacuum state can be defined by using appropriate boundary conditions for the field in the near-horizon region. We want to define two vacuum states by using the two natural notions of time translation of the Kerr geometry, namely, the Killing time and the proper time as measured by a congruence of freely falling observers. To define the {\it out} vacuum state, we consider a set of orthonormal (with respect to the standard Klein-Gordon product) modes that in the near-horizon region are outgoing modes of the form $f^{out}_{\tilde{w},A,m}=f^{out}_{\tilde{w}}(u) Z_{A,m}(\theta,\tilde{\phi})/\sqrt{r^2+a^2}$, where $f^{out}_{\tilde{w}}(u)=e^{-i \tilde{w} u}/\sqrt{4 \pi \tilde{w}}$. We can construct wave packets from the previous modes in the usual way \cite{hawk1} $f^{out}_{j,n}=\frac{1}{\epsilon} \int_{j\epsilon}^{(j+1)\epsilon} d\tilde{w} e^{2\pi i \tilde{w} n/\epsilon} f^{out}_{\tilde{w}}(u)$, with integers $j\ge0$ and $n\gg1$. These packets are localized in the near-horizon region and are peaked around the late time $u=2\pi n/\epsilon$, with width $2\pi/\epsilon$.
Taking $\epsilon$ small ensures that the frequency of the modes is narrowly centered around $\tilde{w}\approx \tilde{w}_j=j\epsilon$. The previous modes have positive norm if $\tilde{w}>0$, and they  are positive frequency modes with respect to the Killing vector $\chi^{\mu}$, which is timelike in the near-horizon region. Because of the potential barrier $V(r)$ of the wave equation, these modes will split up in two parts during their propagation. A fraction $|r_{A,m}|^2$ of the wave packet will be reflected by the potential barrier and will fall down into the black hole, where $r_{A,m}$ is the reflection coefficient of the potential. On the other hand, a fraction $\Gamma_{A,m}=1-|r_{A,m}|^2$ of the wave packet will be transmitted, reaching the asymptotic region where the wave packet is of positive frequency with respect to the standard Killing time $t$.
The modes $f^{out}_{j,n}$ correspond to the so-called {\it up} modes in the eternal extension of the black hole geometry.
By expanding the field operator as $\Phi=\sum_{\tilde{w},A,m} a^{out}_{\tilde{w},A,m} f^{out}_{\tilde{w},A,m} + ... $, the {\it out} vacuum state $|out\rangle$ is defined as the state annihilated by the operators $a^{out}_{\tilde{w},A,m}$. Additionally, we specify that $|out\rangle$ contains no ingoing radiation falling into the black hole. The particular form of the modes describing ingoing radiation will not affect the computations of particle production far from the black hole at late times.

In order to define the ${\it in}$ vacuum state, let us write the Kerr line element in the near-horizon region in terms of Kruskal-like coordinates defined as $U=-\kappa^{-1} e^{-\kappa u}$, $V=\kappa^{-1} e^{\kappa v}$, where $\kappa=(r_+ -r_-)/[2 (r^2_++a^2)]$ is the surface gravity of the black hole horizon. By taking $d\theta=d\tilde{\phi}=0$ (without restricting the physics, we take, for simplicity, coordinate $V\sim 0$) the metric has the simple form $ds^2\approx C (dT^2-dR^2)$, where $U=T-R$, $V=T+R$ and $C$ is a finite constant that, without loss of generality, we take equal to $1$. We can see that the $(T,R)$ part of the metric near the horizon has the form of the Minkowski metric. The interval of time $\Delta T$ then corresponds to the interval of proper time of a radial ($\theta and \tilde{\phi}$ constant) freely falling observer crossing the horizon. Therefore, the modes $f^{in}_{\sigma,A,m}=f^{in}_{\sigma}(T,R) Z_{A,m}(\theta,\tilde{\phi})/\sqrt{r^2+a^2}$, where $f^{in}_{\sigma}(T,R)=e^{-i (\sigma T-K R)}/\sqrt{4 \pi \sigma}$, (with $K^2=\sigma^2-\mu^2$), are positive frequency modes with respect to the freely falling observer for $\sigma>0$. Expanding the field operator as $\Phi=\sum_{\sigma,A,m} a^{in}_{\sigma,A,m} f^{in}_{\sigma,A,m}+... $, the {\it in} vacuum state $|in\rangle$ is defined as the state annihilated by the operators $a^{in}_{\sigma,A,m}$ and containing no ingoing radiation coming from infinity. This state corresponds to the so-called Unruh vacuum defined in the maximally extended geometry.

We can now use Eq. (\ref{eq:new-nord}) to compute the expectation value of the {\it out} number operator in the {\it in} vacuum state. Let us consider a Cauchy hypersurface $\cal{C}$, constructed such that it is a null hypersurface in the near region outside the horizon for $U>U_0$, with $U_0$ a negative constant (we recall that  $U=0$ at the horizon and $U\to-\infty$ when $r\to \infty$). Additionally, for $U\le U_0$ and for $U>0$ (inside the horizon), $\cal{C}$ is a spatial hypersurface. Because we are dealing with sharply localized wave packets near the horizon, the particular value of $U_0$ is unimportant. The integrals in (\ref{eq:new-nord}) are extended over the (null) region of $\cal{C}$ where the wave-packets $f^{out}_{j,n}(u)$ have support. There we have $d\Sigma^{\mu}\partial_{\mu}= du d(cos{\theta})d\tilde{\phi} (r^2+a^2) \  \partial_u$. We can take advantage of the orthogonality of the spheroidal harmonics to integrate the angular part in (\ref{eq:new-nord}). An additional integration by parts gives
\bea \label{bbb}   \langle in|N^{out}_{j,n}|in\rangle &=& 4 \int_{\cal{C}} du_1 du_2  f^{out}_{j,n}(u_1)  f^{out *}_{j,n}(u_2)   \\ \nonumber  & \times& [ \langle in| \partial_{u_1}\Phi_{A,m}(u_1)  \partial_{u_2}\Phi_{A,m}(u_2) |in\rangle \\  &-& \langle out| \partial_{u_1}\Phi_{A,m}(u_1) \partial_{u_2}\Phi_{A,m}(u_2)|out \rangle ]  \nonumber  \, . \eea
The relevant $out$ two-point function in the near-horizon region can be easily computed as a sum in modes $\langle out| \partial_{u_1}\Phi_{A,m}(u_1) \partial_{u_2}\Phi_{A,m}(u_2)|out \rangle=-\frac{1}{4 \pi} \frac{1}{(u_1-u_2)^2}$. The two-point function corresponding to the $|in\rangle$ state can also be computed as a sum in modes, and, when evaluated at the null near-horizon portion of $\cal{C}$, it takes the simple form $ \langle in| \partial_{U_1}\Phi_{A,m}(U_1) \partial_{U_2}\Phi_{A,m}(U_2)|in \rangle=-\frac{1}{4\pi}  \frac{1}{(U_1-U_2)^2}$, where we can explicitly see that, as a by-product of conformal invariance, the mass of the field has disappeared.
Introducing these two-point functions in (\ref{bbb}), we have
\bea \label{cft}
\langle in|N^{out}_{j,n}|in\rangle = 4 \int_{\cal{C}} du_1 du_2 \  f^{out}_{j,n}(u_1)  \ f^{out *}_{j,n}(u_2) \\ \nonumber\left[ -\frac{1}{4 \pi}  \frac{\frac{dU}{du}(u_1) \frac{dU}{du}(u_2)}{(U(u_1)-U(u_2))^2}+\frac{1}{4 \pi}\frac{1}{(u_1-u_2)^2} \right] . \eea
Some important comments are now in order. This expression makes it manifest that the particle number is obtained by comparing the local behavior of the two-point functions of the $2$-dimensional primary fields $\partial_u\Phi_{A,m}(u)$ and $\partial_U\Phi_{A,m}(U)$, together with the conformal transformation $U(u)=-\kappa^{-1}e^{-\kappa u}$. Here, local means the near-horizon region where the wave packets $f^{out}_{j,n}(u)$ have support. Additionally, the conformal transformation $U(u)=-\kappa^{-1}e^{-\kappa u}$ is a local transformation that does not belong to the global group of conformal transformations in two dimensions: $SO(2,2)$. Hence, as explained above, the function between square brackets in (\ref{cft}) is nonvanishing [compare this expression with (\ref{2d2pt})], leading to a nonzero value for $ \langle in|N^{out}_{j,n}|in\rangle$. Additionally, its leading term in an expansion about the coincidence point $u_1\to u_2$ is $-c/(24\pi)\{U,u\}$, where $\{U,u\}$ is the Schwarzian derivative of $U(u)$ and $c=1$ is the central charge of the scalar field. This fact makes manifest the underlying connection between black hole particle production and the conformal anomaly. For wave-packets sharply peaked around $\tilde{w}_j$, simple manipulations lead to
\be \label{final} \langle in|N^{out}_{\tilde{w}_j}|in\rangle  =\frac{-1}{2 \pi \tilde{w}_j} \int dz  e^{-i \tilde{w}_j z} \left[  \frac{(\kappa/2)^2}{\sinh^2{[\frac{\kappa}{2}z]}}-\frac{1}{z^2}   \right]   \nonumber  \ , \ee
where we have defined $z\equiv u_1-u_2$. This integral can be easily computed and results in the well-known planckian distribution of particles. An observer at the late time region of the asymptotic future will observe a thermal flux of particles at temperature $\kappa/2\pi$ with the spectrum additionally modulated by the gray-body coefficients $\Gamma_{A,m}$, $\frac{\Gamma_{A,m}}{e^{2 \pi\tilde{w}/\kappa}-1}$, with $\tilde{w}=w-\Omega_H m$. The scattering coefficients $\Gamma_{A,m}$ can also be obtained by means of the $d=2$ conformal symmetry appearing in the wave equation \cite{hiddencft}.

\noindent{\it Conclusions-} Finite-dimensional $[SO(2,2)]$ conformal invariance of the matter wave equation is related to classical superradiance  \cite{hiddencft}. We have pointed out here that the full local conformal invariance of the wave equation emerging near the horizon accounts for the full quantum phenomena of black hole radiance. This result makes manifest the link between the full radiation spectrum of black holes and conformal symmetry.

\noindent {\it Acknowledgements} This work has been supported by the Spanish Grants FIS2008-06078-C03-02 and CPAN (CSD2007-00042), and an NSF and a UWM RGI grant.\\


\begin{thebibliography}{99}







\bibitem{christensen-fulling} S.M. Christensen and S.A. Fulling, {\it Phys. Rev. D} \textbf{15}, 2088 (1977).

\bibitem{navarro-talavera} J. Navarro-Salas, M. Navarro and C.F. Talavera, {\it Phys. Lett. B} {\bf 356}, 217 (1995). G.Amelino-Camelia and D. Seminara, {\it Class. Quant. Grav.} {\bf 13}, 881 (1996).

\bibitem{wilczek} S.P.Robinson and F.Wilczek,  {\it Phys.Rev.Lett} \textbf{95}, 011303 (2005); S.Iso, H.Umetsu and F.Wilczek, {\it Phys.Rev.Lett.} \textbf{96}, 151302 (2006); {\it Phys.Rev.D} \textbf{74}, 044017 (2006).

\bibitem{fabbri-navarro-salas} A. Fabbri and J. Navarro-Salas, {\it Modeling black hole evaporation}, ICP-World Scientific, London (2005).

\bibitem{agullo-navarro-salas-olmo-parkerPRL1}
I. Agullo, J. Navarro-Salas and G.J. Olmo, {\it Phys. Rev.
Lett.} {\bf 97}, 041302 (2006).

\bibitem{agullo-navarro-salas-olmo-parkerPRD1}I. Agullo et al, {\it Phys. Rev. D}  {\bf 76} 044018 (2007).


\bibitem{iso-morita-umetsu} S.Iso, T.Morita and H.Umetsu, {\it Phys.Rev.D} \textbf{75}, 124004 (2007); L.~Bonora et al.,  Phys.Rev.D {\bf 80} (2009) 084034.

\bibitem{carlip} S.~Carlip, {\it Phys. Rev. Lett}. {\bf 82}, 2828 (1999); {\it Phys. Rew. Lett.} {\bf 99}, 021301 (2007); {\it Gen. Rel. Grav.} {\bf 39}, 1519 (2007).

\bibitem{strominger} A. Strominger, {\it JHEP} {\bf 9802}, 009 (1998).

\bibitem{solodukhin} S.N. Solodukhin, {\it Phys. Lett. B} {\bf 454}, 213 (1999).

\bibitem{kerrcft} M. Guica, T. Hartman, W. Song and A. Strominger, {\it Phys. Rev. D} \textbf{80}, 124008 (2009). A. Castro and F. Larsen, {\it JHEP} \textbf{0912}, 037 (2009)

\bibitem{superradiancekerrcft}  T. Hartman, W. Song and A. Strominger, {\it JHEP} \textbf{1003}, 118 (2010).

\bibitem{hiddencft} A. Castro, A. Maloney and A. Strominger, arxiv:1004:0996v1.

\bibitem{hawk1}
S.W. Hawking, {\it Comm. Math. Phys.} {\bf 43}, 199 (1975).

\bibitem{parker68} L. Parker, {\it Phys. Rev.Lett.} {\bf 21} 562 (1968); {\it Phys. Rev.} {\bf 183}, 1057(1969).


\bibitem{parker-toms}
L. Parker and D. J. Toms, {\it Quantum field theory in curved
spacetime: quantized fields and gravity}, Cambridge University
Press, Cambridge (2009).



\bibitem{difrancesco} P.~Di Francesco, P.~Mathieu and D.~Senechal \emph{Conformal Field
Theory}. Springer, New York(1997).

\bibitem{frolov-novikov} V. P. Frolov and I. D. Novikov, {\it Black hole physics}, Kluwe Academic Publishers, Dordrecht, The Nederlands (1998).

\bibitem{unruh}W.G. Unruh, {\it Phys. Rev.} D {\bf 15}, 365 (1977),  {\it Phys. Rev.} D {\bf 14}, 3251 (1976).


\bibitem{fredenhagen-haag90} K. Fredenhagen and R. Haag, {\it Commun. Math. Phys.} {\bf 127} 273 (1990)


\end{thebibliography}
\end{document}